\documentclass[prl,superscriptaddress,twocolumn,amsmath,aps]{revtex4}
\usepackage{graphicx,subfig} 
\usepackage[utf8]{inputenc}

\renewcommand{\phi}{\varphi}

\begin{document} 
\title{Growing static and dynamic length scales in a glass-forming liquid}

\author{François Sausset}
\email{sausset@lptmc.jussieu.fr}
\affiliation{Institut de Physique Théorique, CEA, CNRS URA 2306, F-91191 Gif sur Yvette, France}

\author{Gilles Tarjus}
\email{tarjus@lptmc.jussieu.fr}
\affiliation{LPTMC, CNRS-UMR7600, Université Pierre et Marie Curie, boîte 121, 4 Place Jussieu, 75252 Paris cedex 05, France}

\date{\today}
\begin{abstract}
We investigate the characteristic length scales associated with the glass transition phenomenon. By studying an atomic glass-forming liquid in negatively curved space, for which the local order is well identified and the amount of frustration opposing the spatial extension of this order is tunable, we provide insight into the structural origin of the main characteristics of the dynamics leading to glass formation. We find that the structural length and the correlation length characterizing the increasing heterogeneity of the dynamics grow together as temperature decreases. However, the system eventually enters a regime in which the former saturates as a result of frustration whereas dynamic correlations keep building up.
\end{abstract}

\maketitle

The spectacular and universal character of the slowdown of relaxation in liquids as one approaches the glass transition seems to call for a detail-independent explanation that involves a growing length scale of one sort or another~\cite{cavagna09,kivelson08}. Indeed, the issue of characteristic length scales has become one of the main themes of most theories of the glass transition. A large body of work has been triggered by the experimental and numerical finding that the dynamics becomes increasingly spatially heterogeneous as it slows down~\cite{ediger00}. Evidence for the existence of a related growing length scale has been obtained either from measures of various kinds of dynamic clustering~\cite{ediger00,donati,weeks00} or from multi-point space-time correlation functions~\cite{dasgupta91,donati,berthier05,berthier07}. On the other hand, despite the rather general proposition that a growing time scale should come with a growing static length scale~\cite{montanari06} and evidence that the dynamical heterogeneities have some structural origin~\cite{widmer04}, the quest for relevant static correlations has been thwarted by obstacles. Among the latter is the well known fact that the simplest pair correlations probed by the static structure factor show no significant features as temperature decreases.

There are several possible, and not mutually excluding, strategies to search for static correlations associated with the glassy properties of a supercooled liquid; such correlations presumably involve some ``hidden'' order parameter that does not show up in simple structural measures such as the static structure factor. One may then think of: (i) studying how far amorphous boundary conditions influence the system, thereby looking at static point-to-set correlations~\cite{bouchaud04,biroli08,montanari06,cavagna09}, (ii) using finite-size analysis for various thermodynamic quantities~\cite{verrochio,karkamar08}, or else (iii), provided that the locally preferred arrangement of the molecules in the liquid has been properly identified~\cite{steinhardt81,dzugutov02,coslovich07,tanaka,anikeenko07,procaccia,wochner09}, investigating the static pair correlations of the associated local order parameter, which amounts to considering multi-particle correlations as in bond-orientational order parameters~\cite{steinhardt81}. In this Letter, we follow the latter route.

We investigate, at a microscopic level, the relationship between structure and dynamics in a glass-forming liquid. We identify  static and dynamic lengths that both grow as relaxation becomes slower and the glass transition is approached. The model we focus on is motivated by the frustration-based approach of the glass transition~\cite{nelson02,tarjus05}: it is a 2-dimensional (2D) monodisperse 6-12 Lennard-Jones liquid whose specificity resides in the negatively curved (hyperbolic) space it is embedded in. This model in ``flat'' (Euclidean) 2D space is not frustrated because the local hexagonal order can freely propagate in space to form a triangular lattice. Placing the system in hyperbolic geometry introduces frustration, whose strength is associated with the curvature of the hyperbolic plane. As a result the liquid does not crystallize and forms a glass upon cooling~\cite{sausset08}. Frustrated hexagonal order in 2D negatively curved space can be thought of as the analog of frustrated icosahedral order in 3D Euclidean space~\cite{nelson02}. We have previously shown that the model displays the common features found in glassformers phenomenology~\cite{sausset08}.


In a nutshell, the hyperbolic plane is a 2D homogeneous space of constant negative curvature $-\kappa^2$ with metric in polar coordinates expressed as $\mathrm{d}^2 s= \mathrm{d}^2r + \left(\sinh(\kappa r)/\kappa \right)^2 \mathrm{d}^2\phi$. The peculiarities of negatively curved space (boundary effects, parallel transport of vectors, absence of global embedding in Euclidean space) entail several tricks for implementing molecular dynamics simulations. For details we refer to~\cite{sausset07,sausset08,saussetforth}. The simulations have been performed in the NVE ensemble at a density $\rho\sigma^2=0.85$ and for a frustration (curvature) parameter $\kappa \sigma$ from $0.02$ to $0.2$ ($\kappa=0$ is the unfrustrated Euclidean plane). To enlarge the system size as needed to compute correlation lengths, we have considered periodic boundary conditions with both octagonal and 14-gonal elementary cells. (In hyperbolic geometry, the elementary cell area $A$ depends on the genus $g$ of the associated compact manifold as $A=4 \pi \kappa ^{-2} (g-1)$, with $g=2$ for the octagon and $3$ for the 14-gone used here~\cite{sausset07}.) Over the range covered by the simulations, we are able to explore three different regimes in the liquid as temperature $T$ decreases~\cite{sausset08}: the ``normal'' liquid range above a temperature $T^*\simeq 0.75$ that is characteristic of hexagonal-like ordering in the Euclidean space, the ``fragile supercooled'' liquid regime, which we have claimed to be controlled by the avoided transition at $T^*$, and a regime dominated by the kinetics of rare irreducible frustration-induced topological defects. The first two regimes presumably represent the experimentally accessible range in fragile glass-forming liquids.

\begin{figure}[tbp]
	\captionsetup[subfloat]{margin=7.1cm}
	\centering
	\subfloat[]{\label{fig:G6} 
		\includegraphics[width=7 cm,trim= 1cm 1.6cm 0.5cm 1cm]{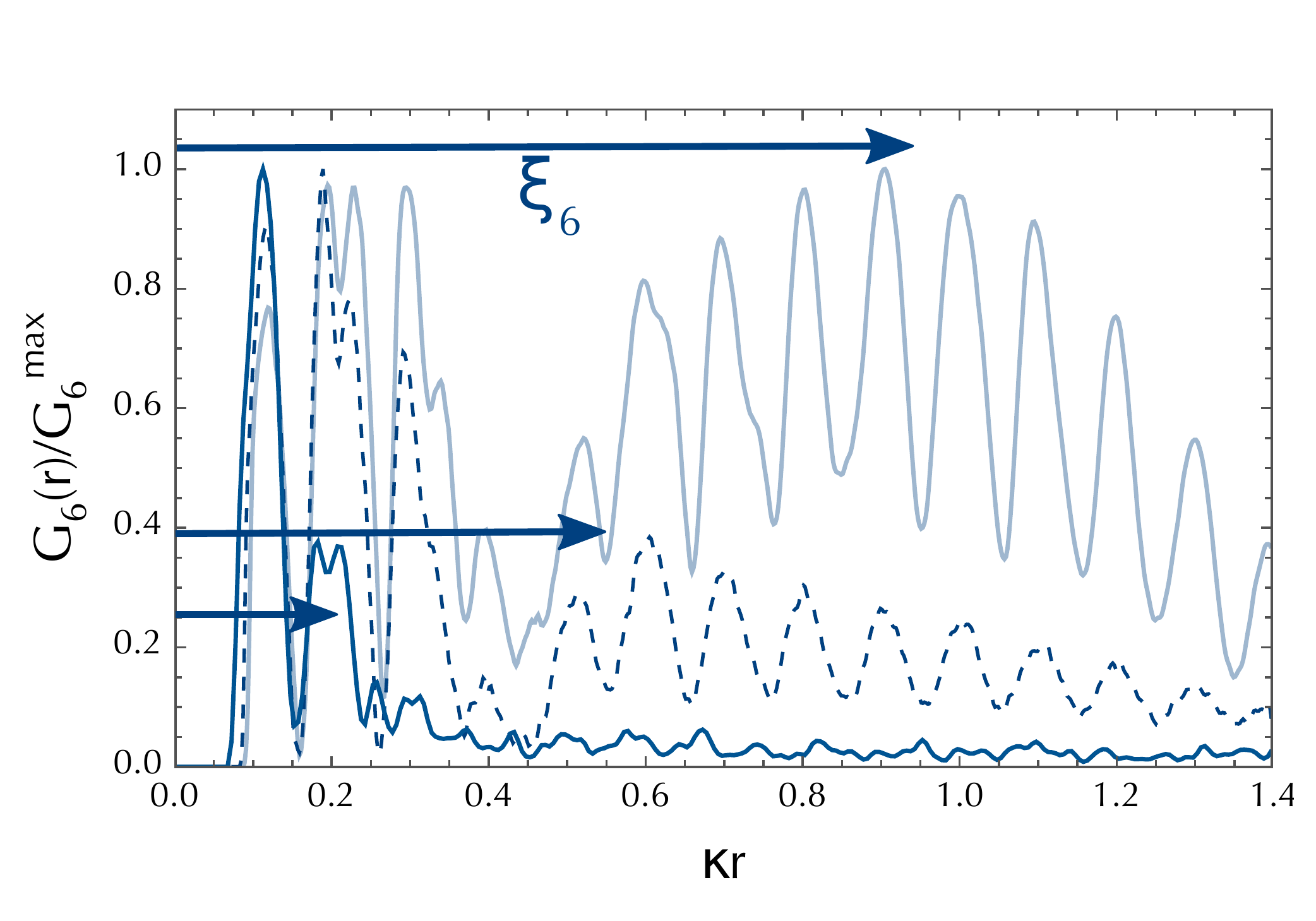}}\\
	\subfloat[]{\label{fig:Xi_6} 
		\includegraphics[width=7 cm,trim= 2cm 1.6cm 0.5cm 1cm]{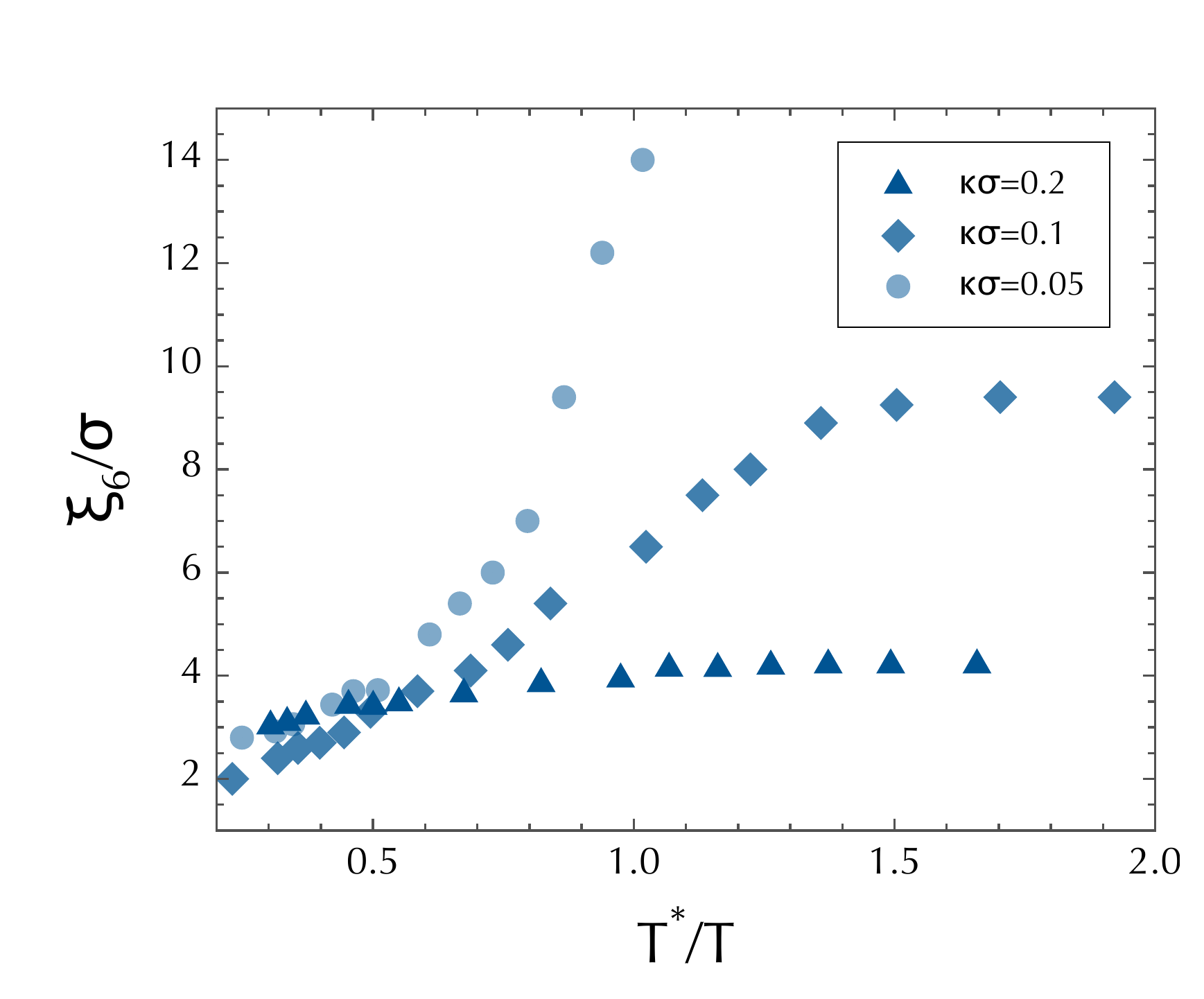}}
\caption{Growth of hexatic local order: (a) $G_6(r)$ normalized by its maximum value for  $\kappa \sigma=0.1$ and $3$   temperatures: $T/T^*=4.35, 1.19, 0.59$ (from bottom to top). (b) Correlation or modulation length $\xi_6$ (extracted either from an exponential fit to the envelope of $G_6(r)$ or from the location of the first modulation peak when present, see (a)) versus $T^*/T$ for $3$ frustrations.}
\label{fig:static_corr}
\end{figure}

Let us first discuss the static structure of the liquid. We have taken advantage of the fact that the locally preferred order is well identified in the present model: it is hexagonal, or more appropriately hexatic, for the small frustrations studied here. In addition to the standard radial pair distribution function $g(r)$, we have then monitored the extension of the local order through two kinds of observables: (i) the topological defects, whose local environment differs from $6$-fold hexagonal arrangement (as obtained through a hyperbolic generalization of the Voronoi tessellation of particle systems), and (ii) the (hexatic) local bond orientational order parameter $\psi_6(\mathbf{j})=(1/N_b)\sum_{<k>}\exp(\mathrm{i}\,6\theta_{jk})$ where the sum is over the $N_b$ nearest neighbors of the $j$th atom and $\theta_{jk}$ is the angle characterizing the ``bond'' between $j$ and $k$. Its pair correlations are obtained from
\begin{equation}
G_6(r) = \frac{1}{Ng(r)}\sum_{i,j=1}^{N}\left\langle \widetilde{\psi}_6(\mathbf{i}|\mathbf{j}) \psi_6(\mathbf{j})^*\right\rangle_{\Gamma_{ij}}\frac{\delta(r_{ij}-r)}{2\pi \kappa^{-1}\sinh( \kappa r)},
\end{equation}
where $\widetilde{\psi}_6(\mathbf{i}|\mathbf{j})$ is the order parameter when parallel transported from point $\mathbf{r}_i$ to point $\mathbf{r}_j$  along the path $\Gamma_{ij}$, chosen here as a geodesic, and $r_{ij}$ is the distance between the two points along $\Gamma_{ij}$; the fraction with the delta function is appropriate for the hyperbolic metric. In addition, we have computed a measure of the fluctuations through $\chi_6 = N (\left\langle \Psi_6 \Psi_6^*\right\rangle -\left\langle \Psi_6\right\rangle \left\langle \Psi_6^*\right\rangle)$, where $\Psi_6 = (1/N) \sum_{i}\widetilde{\psi}_6(\mathbf{i}|\mathbf{0})$ with $\mathbf{0}$ the (arbitrary) origin. The results for $G_6(r)$ at $\kappa \sigma=0.1$ are illustrated in Fig.~\ref{fig:G6}. The envelope of the function decays at high $T$ with a correlation length $\xi_6$ and as $T$ is lowered, it shows a modulation that is associated with the random pattern of frustration-limited hexatic domains (see Fig.~3 in~\cite{sausset08}); as shown in the figure, the characteristic length $\xi_6$ then becomes a modulation length~\cite{nussinov09}. The $T$ dependence of $\xi_6$ is displayed in Fig.~\ref{fig:Xi_6}: $\xi_6$ grows with decreasing $T$ and saturates at a value $\simeq \kappa^{-1}$ which, as seen from the configurations and their Voronoi tessellation~\cite{sausset08}, represents the average distance between the irreducible defects (negative $7$-fold disclinations) that destroy extended hexatic order.  Note that due to the large system sizes required and the very time-consuming numerical procedures, we have not been able to compute the associated correlation length for the longest relaxation times and the lowest curvature (e.g., $\xi_6$ at $\kappa \sigma=0.05$ has not yet saturated in our presently accessible range).

\begin{figure}[tbp]
	\centering
		\includegraphics[width=7 cm,trim= 1cm 1.5cm 0.5cm 1cm]{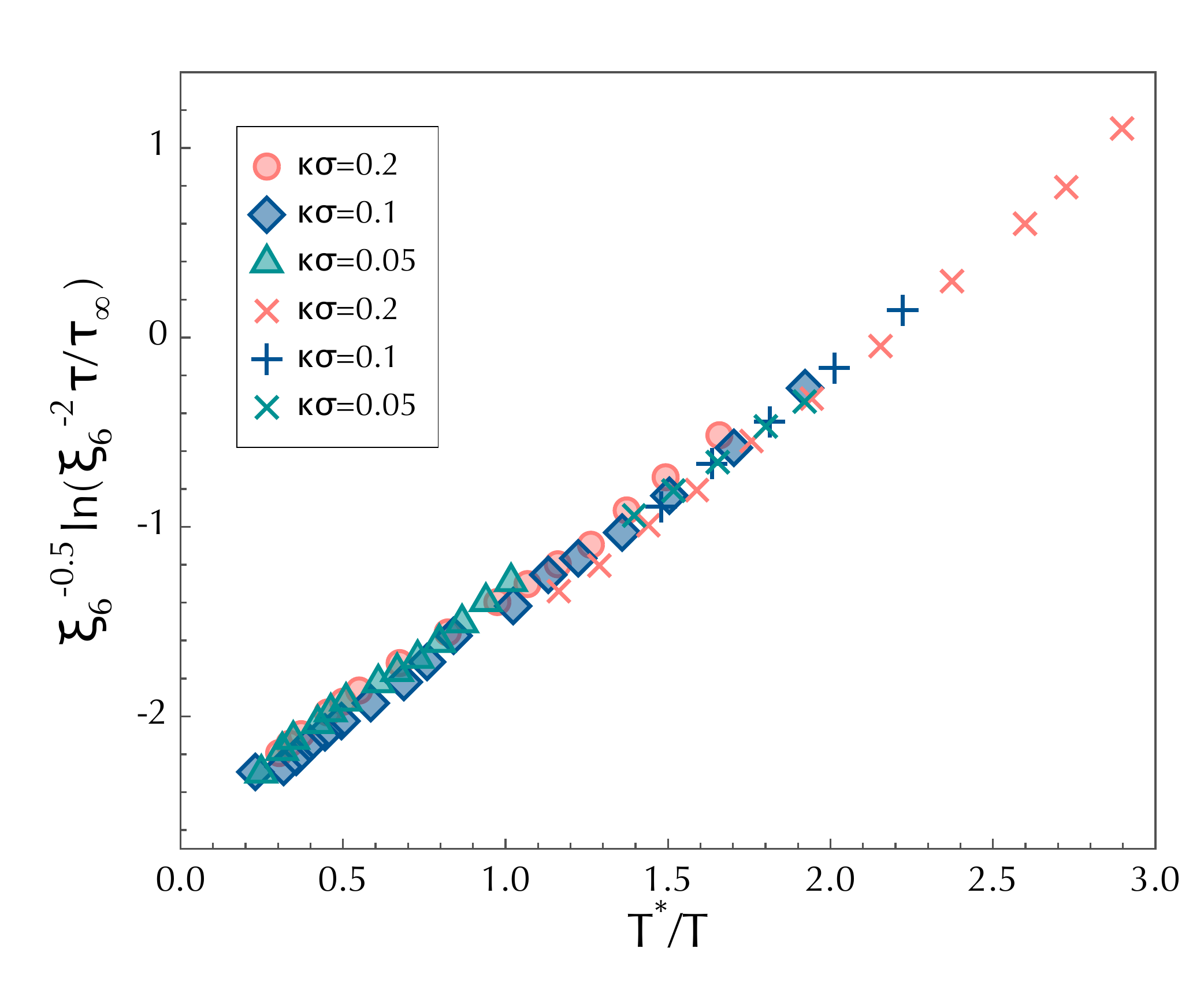}
\caption{Relation between the growth of the relaxation time $\tau$ and that of the static length $\xi_6$: scaled relaxation time $\xi_6^{-0.5} \log(\xi_6^{-2} \, \tau/\tau_{\infty})$ (with $\sigma \equiv 1$) versus $T^*/T$ for the same frustrations as in Fig.~\ref{fig:Xi_6}. The crosses correspond to the regime in which $\xi_6$ has saturated to $\kappa^{-1}$.}
\label{fig:rescaling}
\end{figure}

The connection between spatial extension of the local order and slowdown of the relaxation is shown in Fig.~\ref{fig:rescaling}. From the consideration of defect dynamics at low $T$~\cite{nelson,sausset08} and a scaling hypothesis, we expect that the relaxation time $\tau$ behaves as
\begin{equation}
\tau(T) \simeq \tau_{\infty}\, f\left(\xi_6\right)\, \exp \left( \frac{E(\xi_6)}{T}\right)  , \; \mathrm{with} \; E(\xi_6)\sim \xi_6^{\psi}
\end{equation}
and $0 \leq \psi \leq 1$~\cite{saussetforth}, and $f\left(\xi_6\right) \simeq \xi_6^2\,\exp(-E(\xi_6)/T_x)$ where $T_x$ is a low-$T$ crossover below which hexatic order has saturated (see below). The best fit to the data gives an exponent $\psi \simeq 0.5$ (see Fig.~\ref{fig:rescaling}), but this value is only indicative due to the limited range of data. In connection with the previous observation that the super-Arrhenius character (fragility) increases with decreasing frustration~\cite{sausset08}, this provides strong evidence that in the accessible $T$ range, the slowing down of the dynamics is controlled by the (frustrated) extension of the local  order due to the proximity of the avoided transition at $T^*$~\cite{tarjus05}.

We now turn to the characterization of the spatial correlations in the dynamics. To this end, we characterize the average local dynamics and its fluctuations. From the distance travelled by any atom $j$ during time $t$, $d_j(t)$, we compute the hyperbolic generalization of the local contribution to the ``instantaneous'' self intermediate scattering function, 
$f_{s,j}(k,t) =  P_{-\frac{1}{2}+\mathrm{i}\frac{k}{\kappa}}(\cosh(\kappa d_j(t))$, where $P_{-\frac{1}{2}+\mathrm{i}\frac{k}{\kappa}}$ is a Legendre function of the first kind and $k$ is chosen as usual near the maximum of the static structure factor~\footnote{Note that we focus here on the dynamics probed at a local, atomic, scale. It would be interesting to consider in addition the presumably slower dynamics of the local hexatic order parameter probed on the scale $\xi_6$ and to check along the lines of Ref.~\cite{tanaka2} what are the consequences of a possible additional time scale on the ergodic behavior of the liquid.}. From this, one has access to (i) the average dynamics $F_s(k,t) =  (1/N)\sum_{j=1}^{N} \langle f_{s,j}(k,t) \rangle$, from which we have extracted the relaxation time $\tau$ used above~\cite{sausset08}, (ii) the fluctuations $\chi_4 (t) = (1/N) \langle (\sum_{j=1}^{N}\delta f_{s,j}(k,t))^2\rangle$ with $\delta f_{s,j}=f_{s,j} - \langle f_{s,j}\rangle$, and (iii) the spatial correlations in the dynamics,
\begin{equation}
\mathcal{G}_4(r,t)= \frac{1}{N} \sum_{i,j=1}^{N}\left\langle  \delta f_{s,i}(k,t) \delta f_{s,j}(k,t)\right\rangle \frac{\delta(r_{ij}-r)}{2\pi \kappa^{-1}\sinh( \kappa r)},
\end{equation}
whose integral over the whole system area gives back $\chi_4(t)$.  (As for Eq.~(1), Eq.~(2) reduces to the familiar Euclidean expression when $\kappa \rightarrow 0$.) We have focused on the maximum $\chi_4^{max}$ of $\chi_4 (t)$ and on $\mathcal{G}_4(r,t)$ for the time $t_{max}\sim \tau$ at which this maximum occurs;  from an exponential fit of the envelope of the latter function, we have extracted a dynamic correlation length $\xi_4$.


\begin{figure}[tbp]
	\centering
		\includegraphics[width=7 cm,trim= 1cm 1.5cm 0.5cm 1cm]{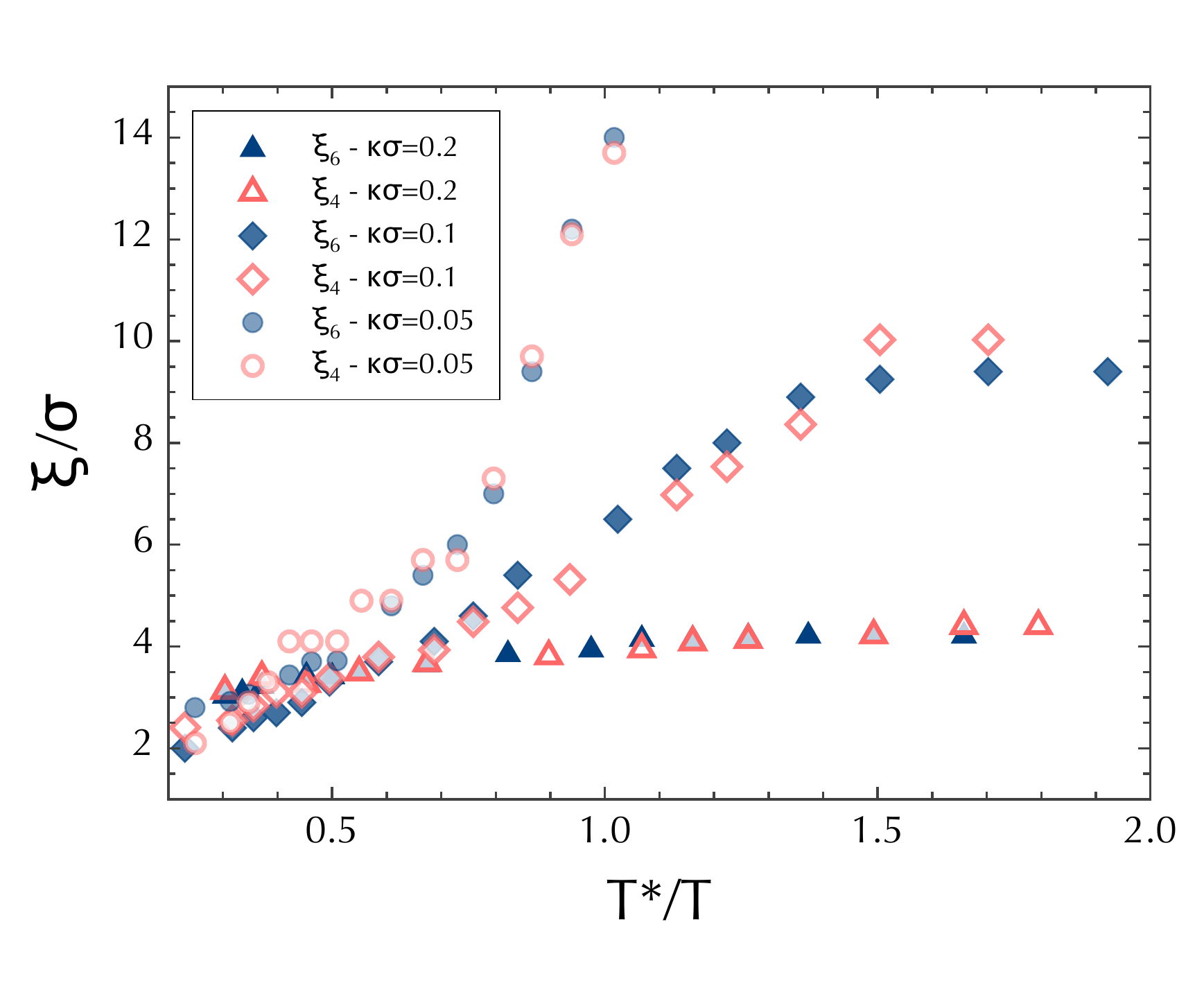}
\caption{Static ($\xi_6$) and dynamic ($\xi_4$) length scales versus $T^*/T$ for $3$ frustrations: from bottom to top, $\kappa \sigma=0.2, 0.1, 0.05$. $\xi_4$ is shifted and divided by a $T$-independent factor.}
\label{fig:lengths}
\end{figure}

A comparison between the ``dynamic'' length scale $\xi_4$ and the static one $\xi_6$ is displayed in Fig.~\ref{fig:lengths}. The $T$ dependences of the two quantities are strikingly similar, with both lengths saturating at low enough $T$ (when accessible) to a value determined by the curvature ($\xi_6$ roughly saturates to $\kappa^{-1}$ and $\xi_4$, which is shifted and divided by a $T$-independent factor in Fig.~\ref{fig:lengths}, to about $0.5 \kappa^{-1}$); $\xi_4$ and $\xi_6$ appear linearly related to each other. This shows that the buildup of spatial correlations in the dynamics, which is associated with growing dynamical heterogeneities, and the one in the static local order parameter, which has been seen above to control the slowdown of relaxation, are directly connected in the $T$ range from normal liquid to the beginning of the regime dominated by irreducible defects.

\begin{figure}[tbp]
	\captionsetup[subfloat]{margin=6cm}
	\centering
	\subfloat[]{\label{fig:Xi_4} 
		\includegraphics[width=7 cm,trim= 1cm 2cm 0.55cm 1cm]{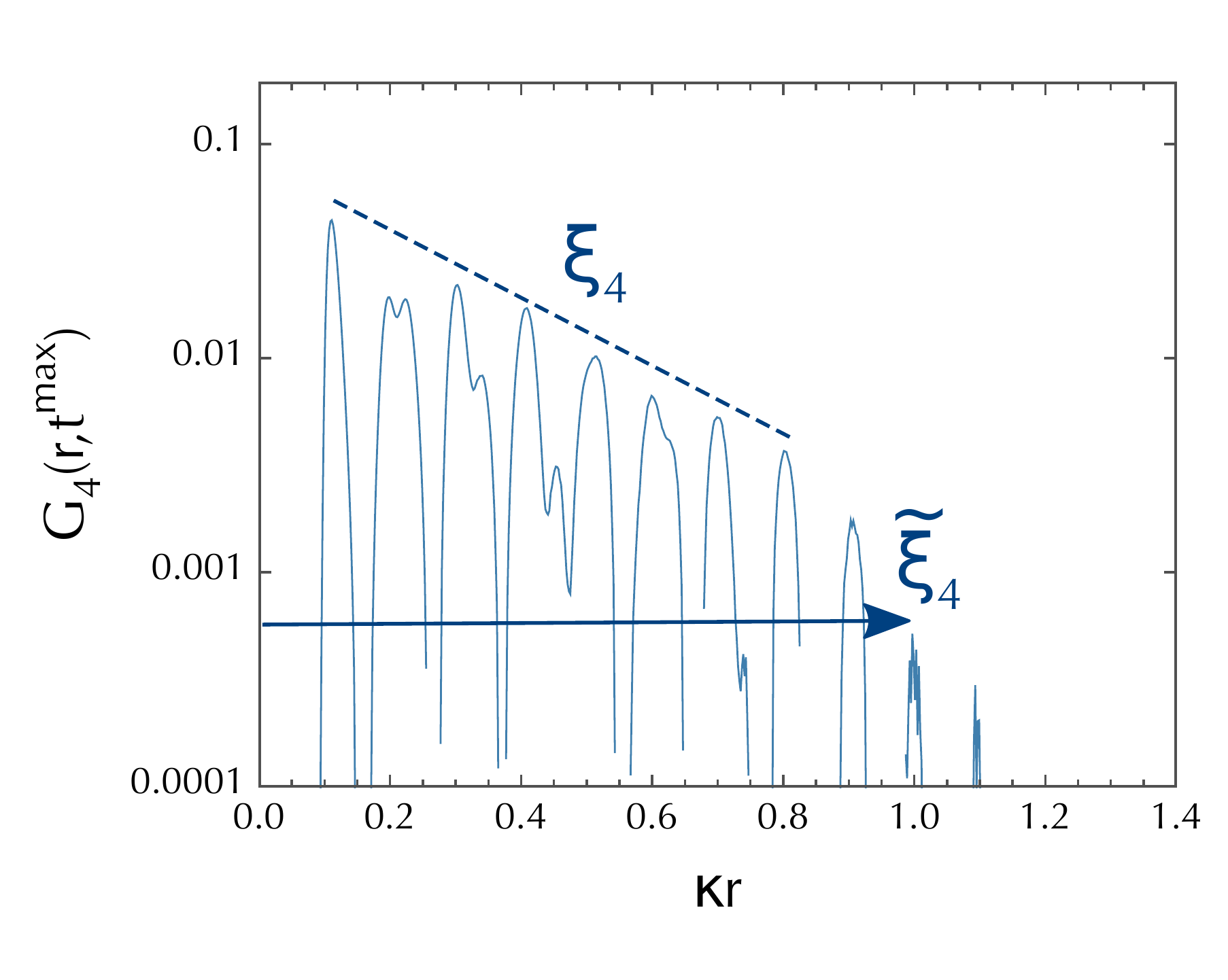}}\\
	\subfloat[]{\label{fig:Chi_4} 
		\includegraphics[width=6 cm,trim= 2cm 1.6cm 1.75cm 1.5cm]{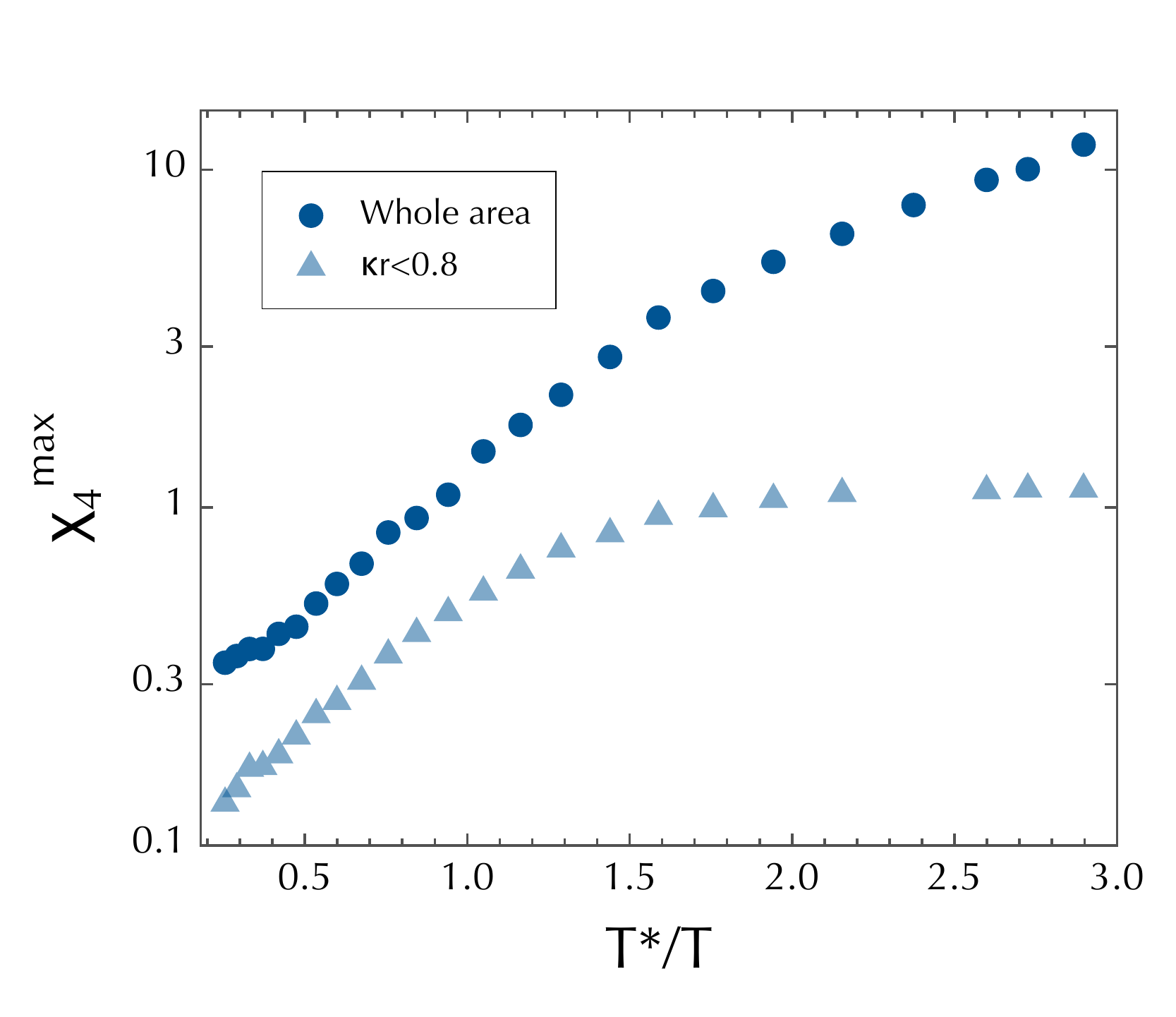}}
\caption{Crossover in the space-time correlations: (a) $\mathcal{G}_4(r,t_{max})$ for $T/T^*=0.82$ and $\kappa \sigma=0.2$; $\xi_4$, characteristic of the exponential decay, is near saturation but an additional length $\widetilde{\xi}_4 > \xi_4$ emerges. (b) For the same conditions, comparison between $\chi_4^{max}$ obtained for the whole system area and by integrating $\mathcal{G}_4(r,t_{max})$ up to $r_c\simeq 0.8 \kappa^{-1}$.}
\label{fig:G4}
\end{figure}

Although we only have limited information on the low $T$ behavior below some crossover $T_x$, once the lengths $\xi_6$ and $\xi_4$ have saturated, it is interesting to push further the analysis of this regime. We anticipate that hexatic order no longer evolves because it has reached its frustration-induced limit. This is confirmed by the behavior of the static ``susceptibility'' $\chi_6(T)$ (not shown here) that saturates just as $\xi_6$ does. On the other hand, the dynamic susceptibility $\chi_4^{max}$ keeps growing with decreasing $T$, albeit with a change of rate: $\chi_4^{max}$ roughly crosses over to a $1/T^2$ behavior (see Fig.~4). Spatial correlations are still increasing as the dynamics slows down in the irreducible-defect regime. A close examination of $\mathcal{G}_4(r,t_{max})$ hints at what may happen: when $\xi_4$ saturates, another length $\widetilde{\xi}_4$ emerges, which in our model corresponds to a  faster than exponential decay of the envelope of $\mathcal{G}_4$ beyond the saturation length (see Fig.~\ref{fig:Xi_4}). Although we cannot follow the growth of this new dynamic length, we can indirectly measure its effect by comparing, as illustrated by Fig.~\ref{fig:Chi_4}, the dynamic susceptibility $\chi_4^{max}$ obtained for the full system area and that computed by truncating the spatial integration of $\mathcal{G}_4(r,t_{max})$ to a cutoff distance $r_c\simeq 0.8 \kappa^{-1}$, i.e. larger than $\xi_4$ at saturation but smaller than $\widetilde{\xi}_4$. The latter saturates as $T$ decreases whereas, as stated before, the former keeps increasing with just a change of rate. We can thus conclude that in the irreducible-defect regime below $T_x$, there is a decoupling between the spatial correlations in the dynamics and in the frustrated hexatic order. To our knowledge this is the first time that such a phenomenon has been observed.

We stress that the crossover to the irreducible-defect regime is likely to be \textit{intrinsic to frustration} rather than a mere peculiarity of hyperbolic space. In any system in which growth of the locally preferred order is frustrated, one expects that there is an intrinsic frustration length at which the spatial extent of the local order saturates~\cite{tarjus05}. For instance, this should be the case in the 2D models of spherical polydisperse particles when polydispersity is higher than the threshold value above which crystallization in hexagonal structures is avoided~\cite{tanaka}, as well as in 3D models with icosahedral local order~\cite{steinhardt81,dzugutov02,coslovich07}. In the temperature range in which saturation has occurred, there is an irreducible density of defects that essentially stays constant with decreasing temperature. If frustration is not too strong, so that the average distance between defects is large, the initial part of this irreducible-defect regime is expected to lead to simple Arrhenius dependence of the relaxation time, with the activation barrier determined by the energy scale of the defects. What happens at still lower $T$ is only subject to speculation. In principle, the defects at least slightly interact and may be subject to kinetic constraints induced by the locally ordered environment.

The present study provides insight into the structural origin of the main characteristics of the dynamics (slowdown of relaxation and increasing heterogeneous character) in a glass-forming liquid. Interestingly, it is found that frustration, a phenomenon which has been hypothesized to be ubiquitous in liquids~\cite{nelson02,tarjus05,tanaka}, imposes a saturation of the spatial extent of the local order leading to a crossover to a low-$T$ regime dominated by rare irreducible topological defects. The slowdown and heterogeneous character of the dynamics then decouple from the frustrated local order. Whether this further results in a simple noncooperative regime or involves a new type of static correlations remains to be investigated. A first step in this direction would be to compute defect-defect correlations as well as point-to-set correlations in the present model.

\end{document}